\documentclass[12pt]{article}
\usepackage[utf8]{inputenc}
\usepackage[english]{babel}
 \usepackage{amsthm}
 \usepackage[flushleft]{threeparttable}

 \usepackage{dsfont}
\usepackage{amsmath}
\usepackage{graphics,graphicx}
\usepackage[comma]{natbib}
\usepackage{tikz}
\usetikzlibrary{arrows,shapes, positioning}
\usepackage{ifthen}
\usepackage{url}
\usepackage{fancyhdr}
\usepackage{amssymb}
\usepackage{graphicx}
\usepackage{color}
\usepackage{colortbl}
\usepackage{adjustbox}
\usepackage[normalem]{ulem}
\usepackage{refcount}

\newcommand{\myfootnotetext}[1]{\footnotetext{#1\label{fn:text}%
        \edef\fnmark{\getpagerefnumber{fn:mark}}%
        \edef\fntext{\getpagerefnumber{fn:text}}%
        \ifx\fnmark\fntext\else\ClassWarning{}{footnote mark and text on different pages!}\fi}}

\newcommand{\blind}{0}

\newcommand{\bsL}{\boldsymbol{\Lambda}}
\newcommand{\bsP}{\boldsymbol{\Psi}}

\newcommand{\bsS}{\boldsymbol{\Sigma}}

\newcommand{\bsp}{\boldsymbol{\Phi}}

\newcommand{\mbf}{\mathbf{f}}

\newcommand{\mbS}{\mathbf{S}}

\newcommand{\mbx}{\mathbf{x}}

\newcommand{\mbX}{\mathbf{X}}

\newcommand{\mbI}{\mathbf{I}}

\newcommand{\mbl}{\mathbf{l}}

\newcommand{\mbQ}{\mathbf{Q}}
\newcommand{\mbN}{\mathbf{N}}
\newcommand{\mbU}{\mathbf{U}}
\newcommand{\mbV}{\mathbf{V}}
\newcommand{\diag}{\rm{diag}}
\newcommand{\mbe}{\mathbf{e}}

\begin{document}

\def\spacingset#1{\renewcommand{\baselinestretch}%
{#1}\small\normalsize} \spacingset{1}

%%%%%%%%%%%%%%%%%%%%%%%%%%%%%%%%%%%%%%%%%%%%%%%%%%%%%%%%%%%%%%%%%%%%%%%%%%%%%%
\if0\blind
{
  \title{\bf Bayesian Multi-study Factor Analysis \\ for High-throughput Biological Data}
  \date{\vspace{-5ex}}
  \author{Roberta De Vito,\\
   \small  Department of Computer Science, Princeton University\\
    Ruggero Bellio, \\
    \small Department of Economics and Statistics, University of Udine \\
      Lorenzo Trippa \\
   \small  Department of Biostatistics and Computational Biology, Dana-Farber Cancer Institute\\
   \small Department of Biostatistics, Harvard T. H. Chan School of Public Health\\
    and \\
    Giovanni Parmigiani \\
    \small Department of Biostatistics and Computational Biology, Dana-Farber Cancer Institute\\
    \small Department of Biostatistics, Harvard T. H. Chan School of Public Health}
  \maketitle
} \fi

\if1\blind
{
  \bigskip
  \bigskip
  \bigskip
  \begin{center}
    {\LARGE\bf Multi-study Factor Analysis}
\end{center}
  \medskip
} \fi

\bigskip

%\begin{comment}
\begin{abstract}

This paper presents a new modeling strategy for joint unsupervised analysis of multiple high-throughput biological studies. As in Multi-study Factor Analysis, our goals are to identify both common factors shared across studies and study-specific factors. Our approach is motivated by the growing body of high-throughput studies in biomedical research, as exemplified by the comprehensive set of expression data on breast tumors considered in our case study. To handle high-dimensional studies, we extend Multi-study Factor Analysis using a Bayesian approach that imposes sparsity. Specifically, we generalize the sparse Bayesian infinite factor model to multiple studies. We also devise novel solutions for the identification of the loading matrices: we recover the loading matrices of interest ex-post, by adapting the orthogonal Procrustes approach. Computationally, we propose an efficient and fast Gibbs sampling approach. Through an extensive simulation analysis, we show that the proposed approach performs very well in a range of different scenarios, and outperforms standard Factor analysis in all the scenarios identifying replicable signal in unsupervised genomic applications. The results of our analysis of breast cancer gene expression across seven studies identified replicable gene patterns, clearly related to well-known breast cancer pathways. An R package is implemented and available on GitHub.

\end{abstract}
%\end{comment}

\noindent%
 \vfill

\newpage
\spacingset{1.45} % DON'T change the spacing!

\section{Introduction}

High-throughput assays are transforming the study of biology, and are generating a rich,
complex and diverse collection of high-dimensional data sets. Joint analyses combining data from different studies and technologies are crucial to improve accuracy of conclusions and to produce generalizable knowledge. 

Most measurements from high-throughput experiments display variation arising
from both biological and artifactual sources. 
Within a study, effects driven by unique issues with the experimental conditions of a specific laboratory or technology can be so large to surpass the biological signal for many biological features \citep{aach2000}. In gene expression, for example, large systematic differences arising from different laboratories or technological platforms have been long recognized \citep{irizarry2003,shi2006,kerr2007}. Systematic collections of gene expression data, collected with technologies that have evolved over time, are widely available, as exemplified by the breast cancer datasets that motivate our work, described in Section~2.

A strength of multi-study analyses is that, generally, genuine biological signal is more likely than spurious signal to be present in multiple studies, particularly when studies are collected from biologically similar populations. Thus, multi-study analyses offer the opportunity to learn replicable features shared among multiple studies.
Discovering these features is, broadly speaking, more valuable than discovering signal in a single study. Joint analyses of multiple genomic datasets have begun more than a decade ago, they are now increasingly common, and can be highly successful \citep{Rhodes:2002ko,Huttenhower:2006wq,Gao:2014tj,Pharoah:2013tm,Riester:2014ga,Ciriello:2013js}. Many such analyses focus on identifying parameters that relate biological features measured at high throughput to phenotypes. These effects can be replicable, though signal extraction across studies can be challenging \citep{Garrett-Mayer2008}. 

An important goal in high-dimensional data analysis is the unsupervised identification of latent components or factors. Despite the importance of this goal, the development of formal statistical approaches for unsupervised multi-study analyses is relatively unexplored. 

In applications, joint unsupervised analyses of high-throughput biological studies often proceed by pooling all the data. Despite their success, 
these studies rely critically on simplified methods of analysis to capture common signal. For example \cite{wang2011} and \cite{edefonti2012} stack  all studies and then perform standard analyses, such as factor analysis (FA) or Principal Component Analysis (PCA).  The results will capture some common features,
but the information about study-specific components will likely be lost, and ignoring it could compromise the accuracy of the common factors found. 

Alternatively, it is also common to analyze each study separately and then heuristically explore common structures from the results \citep{Hayes2006}. 
Co-Inertia Analysis (CIA) \citep{dray2003} explores the common structure of two different sets of variables by first separately performing dimension reduction on each set to estimate factor scores, and then investigating the correlation between these factors. Multiple Co-Inertia Analysis (MCIA) is a generalization of CIA to more than two data sets, which projects different studies into a common hyperspace \citep{meng2014}. Multiple Factor analysis (MFA) \citep{abdi2013} is an extension of PCA  and consists of three steps. The first step applies PCA to each study. In the second step, each data set is normalized by dividing by the first singular value of the covariance matrix. In the third step, these normalized data are stacked by row creating a single data set to which PCA is then applied. 

In practice, there is a need to automatically and rigorously model across studies the common signal that can reliably be identified, while at the same time modeling study-specific variation. 
A methodological tool for this task is Multi-Study Factor Analysis (MSFA), recently introduced in \citet{DV2016}. Inspired by models used in the social sciences, MSFA extends FA to the joint analysis of multiple studies, 
separately estimating signal reproducibly shared across multiple studies 
from study-specific components arising from artifactual and population-specific sources of variation. This dual goal clearly sets MSFA aside from 
earlier applications of FA to gene expression studies, such as 
\citet{carvalho2008}, \citet{friguet2009}, \citet{blum2010}, or \citet{runcie2013}. 

The MSFA methodology in \citet{DV2016} is limited to settings where enough samples are available in each study, and no sparsity is expected or necessary. This is because
model parameters are estimated by maximum likelihood (MLE) and model selection is performed
by standard information criteria.  In high-throughput biology, the sample size routinely exceeds the number of variables, and it is essential to employ regularization through priors or penalties. 

In this paper we introduce a Bayesian generalization of Multi-study factor analysis. 
Bayesian approaches naturally provide helpful regularization, and offer further advantages, discussed later. We leverage the sparse Bayesian infinite factor model, and generalize the multiplicative gamma prior of \citet{bhattacharya2011} to the MSFA setting, to induce sparsity on each loading matrix. We then sample from the posterior distribution via MCMC, without any ex-ante constraints on the loading matrices. This avoids the order dependence induced by the often-used assumption of a lower-triangular form of the loading matrices \citep{geweke1996, lopes2004}, which was employed by the original MSFA proposal. Although useful inferences can be obtained with careful implementation of the constraint, removing it makes the application of FA much simpler and general. We regard this to be an important advantage of our proposal. 

Our prior and parametrization also facilitate inference on the covariance matrices and precision matrices of the observed variables. These are often important goals.  An important example is inference on gene networks, often implemented by first estimating the  covariance matrix through FA \citep{zhao2014, gao2016}. Through the estimation of common factors implied by the decomposition of  the covariance matrix described in \S \ref{sub:def}, the approach we propose allows to detect a common network across the studies, and also to recover the study-specific contributions to gene networks.

The original implementation of the sparse Bayesian infinite factor model \citet{bhattacharya2011} truncates the dimension of the loading matrices at a fixed value. In MSFA, this point is even more important, since our model introduces $(S+1)$ loading matrices if there are $S$ studies. We suggest a pragmatic approach, where the number of dimensions is chosen based on a simple eigenvalue decomposition of covariance matrices obtained as output of the MCMC sampling from the posterior. The specific choice of prior makes the choice of the dimension less critical than would alternative approaches, as we discuss later.

A further strength of our proposal is the recovery of the loading matrices, which are not estimated in \citet{bhattacharya2011}. 
We leverage the recently proposed  Orthogonal Procrustes (OP) method, introduced in \cite{assmann2016}. OP performs an ex-post recovery of the 
estimated loadings by processing the MCMC output, after fitting the model  without any restrictions. The method provides a satisfactory solution to the rotation invariance of FA.
Our results show that the good properties of OP can be generalized to our multiple study setting.  

The plan of the paper is as follows. Section~2 describes the data. Section~3 introduces the Bayesian Multi-study factor analysis (BMSFA) framework, describes our prior, our extension of OP, and our procedure for choosing the number of shared and study-specific factors. Section~4 presents extensive simulation studies, providing evidence on the performance of BMSFA and comparing it with standard methods. We also investigate determining the truncation level for latent 
factors. Section~\ref{sec:CS} applies BMSFA
to the breast cancer data described in Section~2. Section~5 contains a discussion.

%%%%% 2 %%%%%%

\section{The Breast Cancer Data sets}
\label{sub:BCdata}

Breast cancer is both a clinically diverse and a genetically heterogeneous disease \citep{perou2000, planey2016}. The complex nature of breast cancer has been clarified by classifying breast cancer into subtypes using gene expression measurements from tumor samples. Reliably identifying these subtypes has the potential of driving personalized patient treatment regimens \citep{masuda2013} and risk prediction
models \citep{parker2009}. Several groups \citep{sorlie2001, sotiriou2003, hu2006, planey2016} have focused on finding replicable gene expression patterns across different studies, to better classify breast carcinomas into distinct subtypes.

A very valuable statistical approach is unsupervised clustering using different microarrays that query the same set of genes \citep{perou2000, sorlie2001, sorlie2003, castro2016}. A challenge is to characterize the extent to which variation in gene expression, and the resulting subtypes, are stable across different studies \citep{Hayes2006}. When different microarray studies are considered together, one is likely to encounter significant and unknown sources of study-to-study heterogeneity \citep{simon2009, bernau2014}. These sources
include differences in design,  hidden biases, technologies used for
measurements, batch effects, and also variation in the populations studied ---for example, differences in treatment or disease stage and severity. Quantifying these heterogeneities
and dissecting their impact on the replicability of patterns is essential.

A typical bioinformatics analysis pipeline would attempt to remove 
variation attributable to experimental artifacts before further analysis. 
If information on batches of other relevant experimental factors is available, their effects can be addressed \citep{draghici2007systems}.
For example, \cite{sorlie2001} use the SAM (significance analysis of microarrays) algorithm  to detect genes not influenced by batch effect, and then use this set of genes to perform unsupervised cluster analysis. 
In general, it is challenging to fully remove artifactual effects, particularly if they are related to unobserved confounders rather than known batches or factors \citep{draghici2007systems}.

The joint analysis of multiple studies offers the opportunity to understand replicable variation across different studies. The overarching goal of this work is to improve the identification of a stable and replicable signal by simultaneously modeling both the components of variation shared across studies, and those that are study-specific. The latter could include artifacts and batch effects that were not addressed by the study specific preprocessing, as well as biological signal that may hard to replicate or genuinely unique to a study. An example of the latter would be the gene expression signature resulting from the administration of a treatment that is used in one study only.

\begin{table}[t]
\footnotesize
\centering
\begin{tabular}{l c c c c l}
\hline\hline
Study & Adjuvant Therapy & N & N: ER$+$ & 3Q survival &  Reference\\ [0.5ex] % inserts table %heading
\hline
CAL & Chemo, hormonal & 118 & 75 & 42 &  \cite{chin2006} \\
MAINZ & none & 200 & 162 & 120  & \cite{schmidt2008}\\
MSK & combination & 99 & 57 & 76 &  \cite{minn2005}\\
EXPO &  hormonal & 517 & 325 & 126 &  \cite{symmans2010}\\
TRANSBIG &  none & 198 & 134 & 143 &  \cite{desmedt2007}\\
UNT &  none & 133 & 86 & 151 & \cite{sotiriou2006}\\
VDX &  none & 344 & 209 & 44  & \cite{minn2007}\\
\hline
\end{tabular}
\caption{\it The seven data sets considered in the illustration and their characteristics. N is the total number of samples; N: ER$+$ is the number of Estrogen Receptor positive patients. 3Q survival is the third quartile of the survival function for all patients in the study.}
\label{tab: dataset}
\end{table}

In our case study, we consider a systematic collection of publicly available breast cancer microarray studies compiled by \cite{haibe2012}. 
Table~\ref{tab: dataset} provides an overview of the studies, the  corresponding references, sample size, Estrogen Receptor (ER) status prevalence, and survival time. Additional details about these studies, their preprocessing, curation, criteria for inclusion, and public availability are described in \cite{haibe2012}. Four of these studies  only include patients who did not receive hormone therapy or  chemotherapy. Within the Affymetrix technology, genes can be represented by multiple probe-sets. Our analysis considers, for each gene, only the probe-set with maximum mean \citep{miller2011}. As in \cite{bernau2014}, we only consider we only consider genes measured in all the seven studies and focus on the 50\% of genes with higher variance.

%%%%%%%%%%----------------------------------------------------------------------

\section{A Bayesian Framework for multi-study analysis}

This section provides details of our model, in four parts:
\begin{itemize}
\item[i)] Definition of the multi-study factor model sampling distribution;
\item[ii)] Choice of the multiplicative gamma prior \citep{bhattacharya2011}, with shrinkage priors for the loading matrices to incorporate sparsity. Posterior sampling 
 is carried out by Gibbs sampling,  without any constraints on the model parameters;
\item[iii)] Choice of truncation level for the latent 
factor dimensions, determined by a suitable singular value decomposition;
\item[iv)] Recovery of the loading matrices, performed by the OP approach. 
\end{itemize}

\subsection{Model definition}
\label{sub:def}
We consider $S$ studies, each with the same $P$ genomic variables. 
Study $s$, $s=1, \dots, S$, has $n_s$ subjects and
 $P$-dimensional data vector $\mbx_{is}$, $i=1,\ldots,n_s$,  centered at its sample mean.   Our sampling distribution follows the multi-study factor model \citep{DV2016}. The variables in study $s$ are decomposed into $K$ factors shared among all studies, and 
$J_s$ further factors specific to study $s$, as follows:
\begin{equation}
\mbx_{is} = \bsp \mbf_{is} + \bsL_s \mbl_{is} + \mbe_{is} \, .
\label{eqn:MFA}
\end{equation}
Here $\mbf_{is} \sim N_k(\mathbf{0}, \mathbf{I}_{k})$ are the \textit{shared} latent factors, $\bsp$ is their $P \times K$ loading matrix;  $\mbl_{is} \sim N_{j_s}(\mathbf{0}, \mathbf{I}_{j_s})$ are the \textit{study-specific} latent factors and
$\bsL_{s},\; s=1,\dots, S$ are the corresponding $P \times J_s$ loading matrices; lastly, $\mbe_{is}$ is  the $p \times 1$ Gaussian error vector with covariance $\mathbf{\bsP}_{s} = \diag (\psi_{\textit{s}_1}^2, \dots, \psi_{\textit{s}_p}^2)$. 
The resulting marginal distribution of $\mathbf{x}_{is}$ is a multivariate normal with mean vector $\mathbf{0}$ and covariance matrix 
 $\bsS_s = \bsp \bsp^\top+\bsL_s\bsL_s^\top+\bsP_s$. 
The covariance matrix of study $s$ can be rewritten as
\begin{equation}
\bsS_s = \bsS_{\Phi}+ \bsS_{\Lambda_s}+\bsP_s,
 \label{eqn:sigma_mfa}
\end{equation}
where $\bsS_{\Phi}= \bsp \bsp^\top$ is the covariance of the shared factors,  and $\bsS_{\Lambda_s} = \bsL_s\bsL_s^\top$ is the 
covariance of the study-specific factors.  A straightforward implication of (\ref{eqn:sigma_mfa}) is
that $\bsS_{\Phi}$ and $\bsS_{\Lambda_s}$ describe the variability of the $P$ variables in study $s$
that can be interpreted as shared across studies and specific to study $s$, respectively.

The decomposition of $\bsS_s$ is not unique, as there are infinite possibilities to represent it because $\bsp^* = \bsp\mbQ$ and $\bsL_s^* = \bsL_s \mbQ_s$ both satisfy (\ref{eqn:sigma_mfa}) for any two orthogonal matrices $\mbQ$ and $\mbQ_s$.  MSFA  identifies the parameters by imposing constraints on the two factor loadings matrices, such as the lower triangular constraint used in Factor Analysis (FA) \citep{geweke1996, lopes2004} . 
This constraint generates an order dependence among the variables. Thus, 
as noted by \citet{carvalho2008}, 
the choice of the first $K+J_S$ variables becomes an important modeling choice. 

Several approaches focus on the estimation of covariance matrix \citep{bhattacharya2011} or precision matrix \citep{gao2014, zhao2014}. These methods do not require identifiability of the loading matrix. Our approach is also based on this concept: we focus on the estimation of the common variation $\bsS_{\Phi}$ shared among the studies and the variation  specific to each study $\bsS_{\Lambda_s}$. The two matrices $\bsS_{\Phi}$ and $\bsS_{\Lambda_s}$ are only assumed to be positive semidefinite normal matrices, i.e. symmetric matrices with a subset of positive non-null eigenvalues. 

\subsection{ The multiplicative gamma shrinkage prior}
\label{sub:sbif}
We adapt a shrinkage prior from \cite{bhattacharya2011} for both the common and study-specific factor loadings.  
The shrinkage priors favor sparsity by removing some entries of the loading matrix. 
When an element is close to zero, the variable corresponding to the row does not contribute to the common or study-specific latent factor corresponding to the column. In the genomic context, this sparsity 
models the biological reality that only a subset of the genes represented in a cell's transcriptome is participating in a specific biological function  \citep{tegner2003}.
Another important property of the \cite{bhattacharya2011} prior is that the shrinkage towards zero increasing with the column index of the loading matrix. 

Our extension of the multiplicative gamma shrinkage prior to the multiple study setting is as follows.
The prior for the elements of the shared factor loading matrix $\bsp$
is
$$
\phi_{pk} \mid \omega_{pk}, \tau_{k} \sim  N(0,\omega^{-1}_{pk} \tau^{-1}_k), \quad p=1,\dots,P, \; k=1,\dots, \infty,
$$
$$
\omega_{pk} \sim 
\Gamma \left( \frac{\nu}{2},\frac{\nu}{2} \right)
 \;\;\;\;\;\; \tau_{k} = \prod_{l=1}^k \delta_l \;\;\;\;\;\; \delta_1 \sim  \Gamma (a_1,1)  \;\;\;\;\;\; \delta_l \sim  \Gamma (a_2,1),\;\;\;l \geq 2
 $$
where $\delta_l \; (l=1,2,\dots)$ are independent, $\tau_k$ is the global shrinkage parameter for the $k$-th column and $\omega_{pk}$ is the local shrinkage for the element $p$ in column $k$.
We then replicate this scheme to specify  
the prior for the elements of the study-specific factor  loading matrix $\bsL_s$:
$$
\lambda_{pj_s} \mid \omega^s_{pj_s}, \tau^s_{j_s} \sim  N(0,\omega^{s^{-1}}_{pj_s} \tau^{s^{-1}}_{j_s}), \quad p=1,\dots,P, \; j_s=1,\dots, \infty, \; \mbox{ and } s=1,\dots,S,
$$
$$
\omega_{pj_s} \sim 
\Gamma \left( \frac{\nu^s}{2},\frac{\nu^s}{2} \right)
 \;\;\;\;\;\; \tau^s_{j_s} = \prod_{l=1}^{j_s} \delta^s_l \;\;\;\;\;\; \delta^s_1 \sim  \Gamma (a^s_1,1)  \;\;\;\;\;\; \delta^s_l \sim  \Gamma (a^s_2,1),\;\;\;l \geq 2
$$
where $\delta^s_l (l=1,2, \dots)$ are independent, $\tau^s_{j_s}$ is the global shrinkage parameter for the $j_s$ column and $\omega^s_{pj_s}$ is the local shrinkage for the element $p$ in column $j_s$. 

For each of the error variances $\psi_{ps},\,\,p =1,\dots,P$ we assume an inverse
gamma prior
$
\psi_{ps}^{-1} \sim \Gamma (a_{\psi},b_{\psi}).
$
This choice, made also by \citet{bhattacharya2011}, is common
in  standard FA \citep{lopes2004, gao2013, rovckova2016}.
Sampling from the posterior distribution of the model parameters
is carried out by Gibbs sampling. Details are in Supplementary Materials.

\subsection{Choosing the number of latent factors}
\label{sub:nfac}

In practical applications, the number of important latent factors is likely to be small compared to the number of variables
$P$. As suggested by \cite{bhattacharya2011}, the effective number of factors would be small when data are sparse. 
Our approach circumvents the need for pre-specifying the latent dimension since the shrinkage prior gives positive mass to an infinite number of them. However, we need a proper computational strategy for choosing accurate truncation levels $K$ and $J_s$, $s=1,\ldots,S$. Ideally, we would like to retain the relevant factors discarding the redundant ones.

An analogous task for FA is addressed in \citet{bhattacharya2011} who truncate the number of factors to a finite value, usually far smaller than the number of variables $P$.  
This truncation level is chosen by checking the columns of the estimated loading matrix, to assess which ones are formed entirely by elements of negligible size. The fact that the shrinkage implied by the prior increases in later columns greatly simplifies
 this task, compared to  what required by alternative shrinkage priors such as the spike and slab~\citep{carvalho2008}. We use the same idea, though computational details differ. 
 
Our practical method to assess the numbers of shared factors $K$ and study-specific factors $J_s$ is based on singular value decomposition (SVD) and proceeds as follows.
Starting from a considerable number of shared and study-specific factors, we seek  $K \ll P$ and $J_s \ll P$. In the MSFA model, this implies that 
the two matrices $\bsS_{\Phi}$ and $\bsS_{\Lambda_s}$ are singular, with ranks $K$ and $J_s$, respectively. Since these matrices are symmetric, they have $K$ and $J_s$ non-null eigenvalues. 
Based on this,  we compute the eigenvalues $\nu_1, \ldots, \nu_P$ of $\widehat{\bsS}_{\Phi}$, with $\nu_p \geq 0$, $p=1,\ldots,P$, 
ordered in decreasing size. We then choose $K$ as the number of eigenvalues 
larger than a pre-specified  positive threshold, to achieve 
$
\mbU  \, \mbN_K  \mbU^\top \doteq \widehat{\bsS}_{\Phi}  \, ,
$
where $\mbN_K={\rm diag}(\nu_1, \ldots, \nu_K)$, and the columns of $\mbU$, of size $P \times K$,  are given by  $K$ (normalized) eigenvectors of $\widehat{\bsS}_{\Phi}$. We proceed in the same way for $J_s$, $s=1,\ldots,S$. 

\subsection{Recovering loading matrices}
\label{sub:iden}

The method of \S \ref{sub:def}-\ref{sub:nfac} provides a practical route to the estimation of $\bsS_{\Phi}$  and $\bsS_{\Lambda_s}$, but in many applications recovery of the loading matrices is also useful. 
Recently \cite{assmann2016} solved the identification issue in the context of FA by first generating an MCMC sample without any constraints, and then filtering out the possible effect of orthogonal rotations. They solve an Orthogonal Procrustes (OP) problem \citep{gower2004} by building a sequence of  orthogonal matrices defined from the MCMC output. 

Here we extend this procedure to BMSFA.
When the model parameters are not constrained, the Gibbs sampler is said to be orthogonally mixed \citep{assmann2016}, as each chain may produce
different orthogonal transformations (represented by the matrices  $\mbQ$ and $\mbQ_s$) for the factor loadings $\bsp^*$ and $\bsL_s^*$. Starting from 
 a sequence of $R$ draws from the posterior distribution of $\bsp (\bsp^1, \dots, \bsp^R)$, the OP algorithm circumvents this problem by estimation the loading matrices via the following constrained optimization:
\begin{equation}
\left\{\left\{ \tilde\mbQ \right\}_{r=1}^{R}, \tilde\bsp^*\right\}  = \underset{\mbQ^{(r)}, \bsp^*}{\mbox{argmin }}  \sum_{r=1}^{R} L_Q \left( \bsp^*, \bsp^{(r)} \mbQ^{(r)} \right) \,\, \mbox{ s.t. } \, \mbQ^{(r)} \mbQ^{(r)^\top} = \mbI_K, \, r=1, \dots,R 
 \label{eqn:qr}
\end{equation}
where $L_Q$ is the loss function
\begin{equation*}
 L_Q \left( \bsp^*, \bsp^{(r)} \mbQ^{(r)} \right)  = {\rm tr} \left\{ \left( \bsp^{(r)} \mbQ^{(r)} - \bsp^* \right)^\top \left( \bsp^{(r)} \mbQ^{(r)} - \bsp^* \right) \right\}.
 \label{eqn:L_q}
\end{equation*}

The optimization is carried out by iterating two steps:
\begin{enumerate}
    \item Minimize equation (\ref{eqn:qr}), for a given $\bsp^*$ by computing the SVD of $\bsS_{\Phi^*} = \bsp^{(r)} \bsp^{*\top}$ and setting $ \tilde{\mbQ}^{(r)} = \mbU_r \mbV_r$, where $ \mbU_r$ and $\mbV_r$ are the two orthogonal matrices obtained by the SVD at MCMC iteration $r$,
\item Compute
$
 \tilde{\bsp}^{*(r)} = \frac{1}{R} \sum_{r=1}^{R} \bsp^{(r)}  \tilde{\mbQ}^{(r)}.
$
\end{enumerate}
The algorithm is then iterated using the updated value of $\tilde{\bsp}^{*}$ in place of $\bsp^*$. The search stops when subsequent estimates of $\bsp$ are close enough.

This algorithm requires a starting value for $\bsp^*$. \cite{assmann2016} suggests the last iteration of the Gibbs sampler as initial value for $\bsp^*$. The same
procedure can be applied to each of the study-specific loading matrices.
This algorithm provides an approximate solution to identifiability, since  the posterior distribution of the loading matrices is only known in approximate form. Yet, \cite{assmann2016}
show that it can be quite effective. 

The OP procedure is iterative in nature.  However, we verified that typically the first iteration is sufficient
to get close to the final estimate. Since the OP algorithm is computationally
demanding, the one-step version is recommendable. 
All the results of this paper
have been obtained with a single iteration of the OP algorithm.

This point will be further examined for our setting in the following section. 

%%%%%%%%%%%%%%%%%%%%%%%%%%%%%%%%%%%%%%%%%%%%%%%%%%%%%%%%%%%%%%%%%

\section{Simulation Results}
\label{sub:simulation}

\begin{figure}[!b]
	\centering
	\includegraphics[width=15cm]{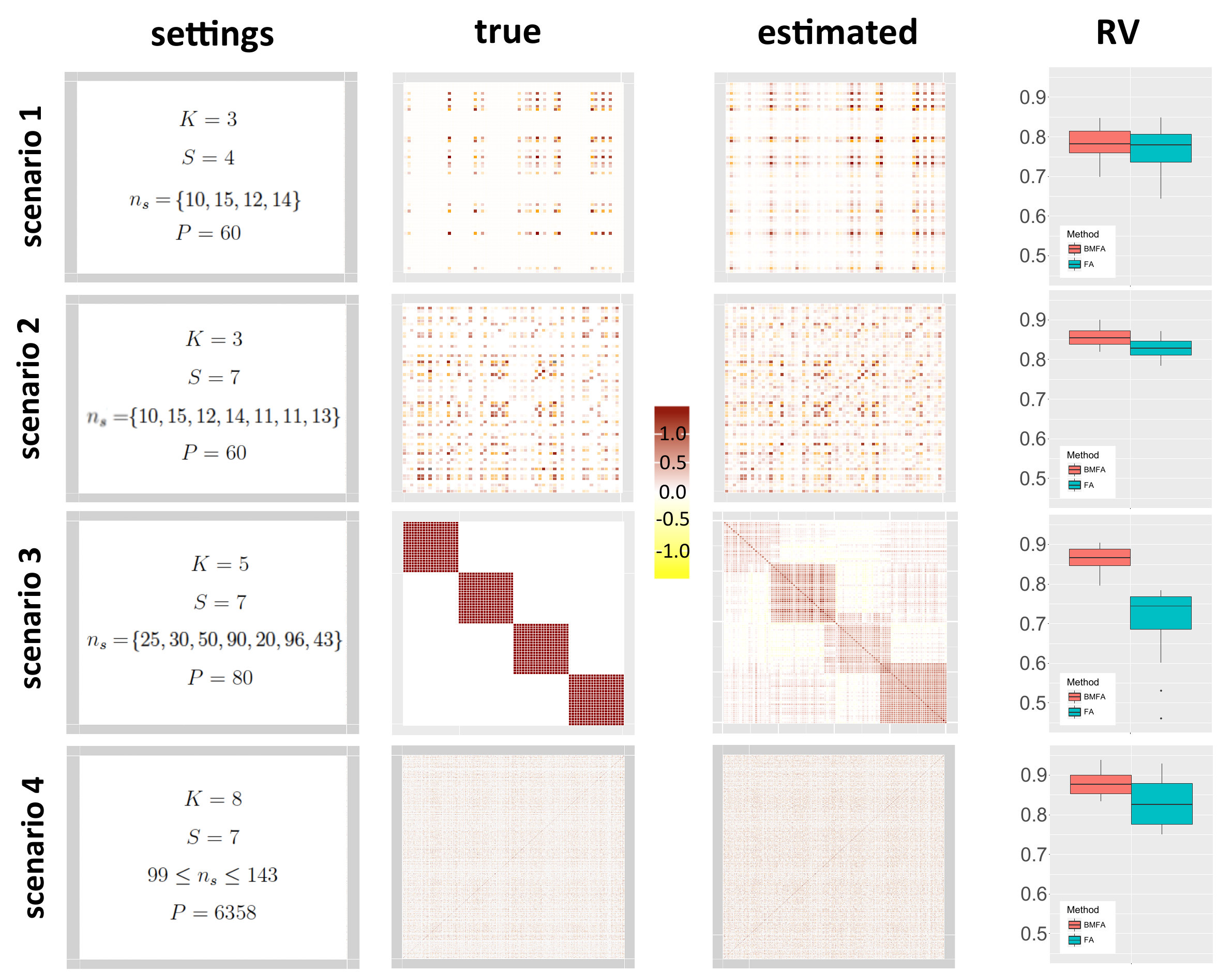}
	\caption{\it \small Covariance matrices $\bsS_{\Phi}$ and their Bayesian estimates in four simulation scenarios. The right column shows the boxplots of RV coefficient between the true and the estimated $\bsS_{\Phi}$.}
	\label{fig: sig}
\end{figure}

In this section we use simulation experiments to assess BMSFA's ability to recover common and study-specific latent dimensions, by itself and in comparison to standard FA applied to the merged datasets. 
We generate 50 datasets from the distributions specified in Table~\ref{tab:simX}. We fixed  $\bsp$, $\bsL_s$ and $\bsP_s$ and thus $\bsS_s$.
\begin{table}[!t]
	\scriptsize
	\parbox{.45\linewidth}{
	\begin{tabular}{c}
		\hline \hline
		$\mbX_s \sim \mbox{MVN}\left( \mathbf{0}, \bsS_s  \right)$ \\
		$\bsS_s = \bsp \bsp^\top+\bsL_s\bsL_s^\top+\bsP_s$\\
		fixed $\bsp$ and $\bsL_s$: sparse matrices with $\approx$ 80 \% of zeros \\
		fixed $\bsp$ and $\bsL_s$: non zero elements drawn once from U$ (-1,1)$\\
		fixed $\bsP_s$: diagonal elements drawn once from U$(0,1)$\\
		\hline \hline
	\end{tabular}
	\caption{\it \small Distributions used to generate observations in study $s$, for simulation experiments. }
\label{tab:simX}
	}
	\hfill \parbox{.45\linewidth}{\begin{tabular}{c}
		\hline        \hline  
		Common factor loadings : $\omega_{pk} \sim \Gamma\left(\frac{\nu =3}{2}, \frac{\nu =3}{2}\right)$  \\
		Study-Specific factor loadings : $\omega_{pj_s} \sim \Gamma\left(\frac{\nu^s =3}{2}, \frac{\nu^s =3}{2}\right)$\\    
		$\delta_1 \sim \Gamma(a_1 = 2.1,1)$ and $\delta_l \sim \Gamma(a_2 = 3.1,1)$ with $l\geq 2$\\
		$\delta_1^s \sim \Gamma(a_1^s = 2.1,1)$ and $\delta_l^s \sim \Gamma(a_2^s = 3.1,1)$ with $l\geq 2$\\
		$\bsP_s^{-1} \sim \Gamma ( a_\psi = 1, b_\psi = 0.3)$\\
		\hline \hline
	\end{tabular}
	\caption{\it \small Prior distributions used in the simulation experiments and real data analysis.}
\label{tab:priSim}
}
\end{table}
We consider four scenarios differing in the number of studies, study sample sizes, and covariance structure (see Figure~\ref{fig: sig}). Scenarios~1 and~2 are similar to \cite{zhao2014}: $n_s$ is chosen to be smaller than $P$ to mimic large $P$ and small $n_s$ conditions while operating with a manageable set of variables for visualization and summarization.  In the Scenario~3 we wish to model a situation where not all the studies have $P\gg n$. Moreover, in this scenario, study-specific factor loadings are large. The motivation behind this scenario is to investigate if our method recovers the shared biological signal in the presence of large study-specific or batch effects, and if it can isolate these sources. In Scenario~4 we closely mimic the data in Table~\ref{tab: dataset}, choosing $S=7$ and matching the sample sizes to those of Table~\ref{tab: dataset}. 
Moreover, in Scenarios 1, 2 and 4 we randomly allocate the zeros in each column of $\bsp$ and $\bsL_S$ (Table~\ref{tab:simX}), while in Scenario 3, we allocate zeros matching the central panel in the third row of Figure~\ref{fig: sig}.
We run the Gibbs sampler for 15000 iterations with a burn-in of 5000 iterations. We set priors as in Table~\ref{tab:priSim}.

We first evaluate, for fixed latent dimension $K$ and $J_s$, BMSFA's ability
to recover the covariance component $\bsS_\Phi$ determined by the shared factors, as well as the shared factors' loadings $\bsp$.   For one randomly selected simulation dataset, 
Figure~\ref{fig: sig} compares the true and estimated elements of $\bsS_\Phi$.  We also present a summary of the analyses of 50 datasets.
To quantify the similarity between $\bsS_\Phi^{true}$ and $\widehat{\bsS}_\Phi$ we use the RV coefficient (Robert and Escouffer, 1976)\cite{} of similarity of  two $P \times P$ matrices $\bsS_1$ and $\bsS_2$:
$$
RV (\mbS_1, \mbS_2) = \frac{tr ((\bsS_1 \bsS_2^\top) (\bsS_1 \bsS_2^\top)}{tr(\bsS_1 \bsS_1^\top)^2 tr (\bsS_2 \bsS_2^\top)^2}.
$$
RV varies in $[0, 1]$. The closer RV is to 1 the more similar the two matrices are.
\citet{smilde2008} argue that the RV coefficient can overestimate similarity between data sets 
in high-dimensions, and propose a modified version that addresses this problem. We use it in Scenario~4, though differences will not be pronounced. The red boxplots in the right column of Figure~\ref{fig: sig} show the RV distributions across 50 simulations in our four scenarios. 

\begin{figure}[!h]
	\centering
	\resizebox{1\textwidth }{!}{ %
		\includegraphics[width=0.3cm]{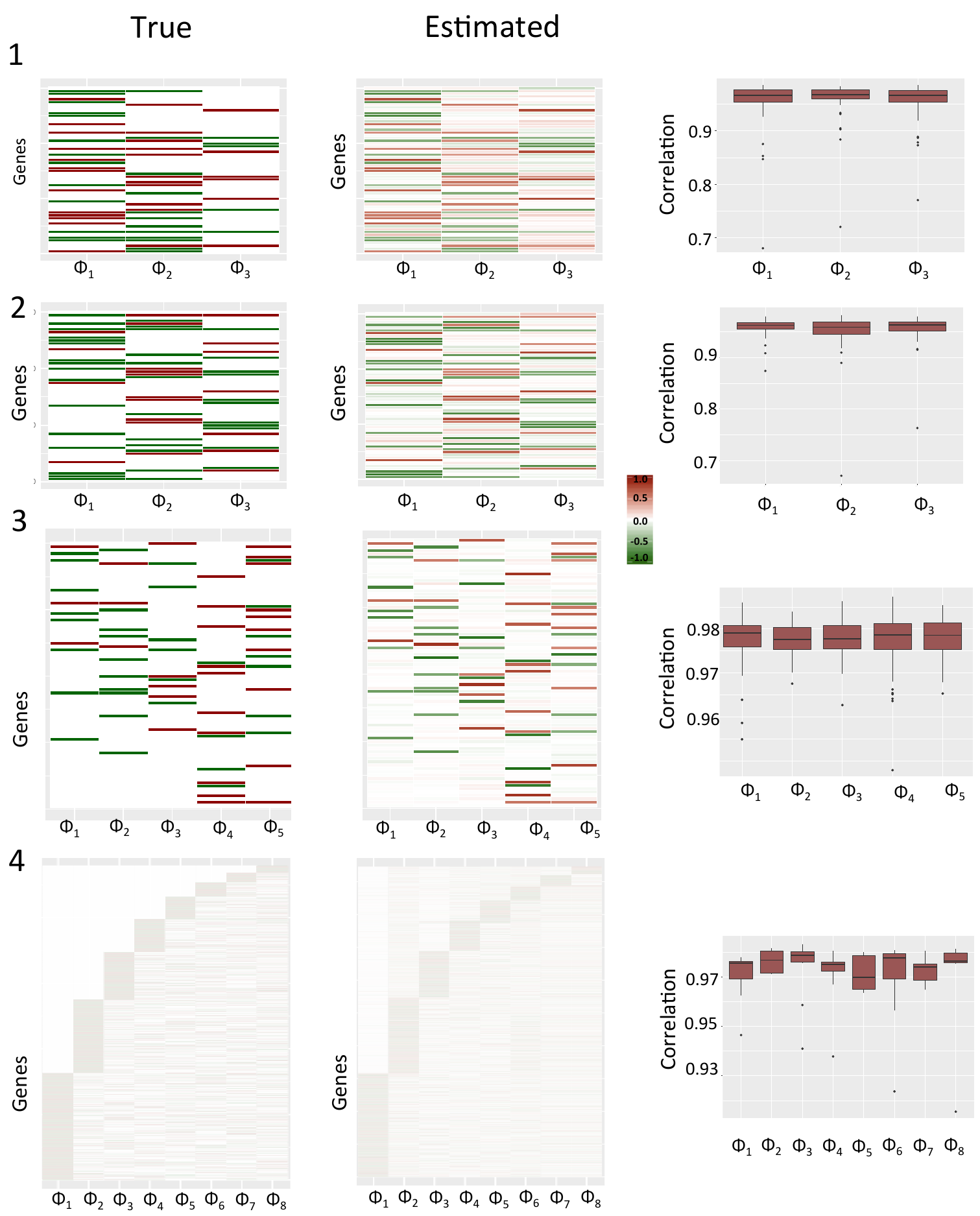}
	}
	\caption{\it Heatmap of the true (left) and estimated (center) shared factor loadings $\bsp$ in the four scenarios of Figure~\ref{fig: sig}. In Scenario 4 we only show common factor loadings $\geq 0.5$. The right column displays boxplots of correlations between the true and estimated common factor loadings over 50 datasets for each scenario.}
	\label{fig: phipro}
\end{figure}

Figure~\ref{fig: phipro} presents a similar analysis comparing the true factor loadings to their estimates obtained through posterior sampling and the OP procedure.  The correlations between true and estimated values in both Figures~\ref{fig: sig}~and~\ref{fig: phipro}) are very high, suggesting that our estimands are well identified and our sampling approaches are appropriate.

Next we compare BMSFA to a Bayesian FA, using the same prior distribution. For Bayesian FA, we stacked all studies into a single dataset, ignoring that samples originate from distinct studies. The RV coefficients for BMSFA are systematically greater than FA's (Figure~\ref{fig: sig}, right column), demonstrating that BMSFA recovers shared factors better than a merged analysis. In Scenario~3 the gap is more pronounced, as study-specific factor loadings are large. In most simulations, FA captures study-specific effects that are not actually shared. BMSFA recovers the shared signal better. Also, the distribution of BMSFA's RV coefficient is narrower than FA's. This comparison illustrates that BMSFA identifies the shared signal  across the studies and improves its estimation compared to standard Bayesian FA.  Moreover, the BMSFA estimations are more efficient compared to the FA estimation, due to the beneficial effects of removing the study-specific components that lack cross-study reproducibility. 

So far we took $K$, the number of shared factors, and $J_s$'s, the numbers of study-specific factors, to be known. We next focus on the latent dimensions calculated via SVD of matrices $\bsS_\Phi$ and $\bsS_{\Lambda_s}$, as described earlier, and using an eigenvalue threshold of 0.05. The simple adaptive method described in \S \ref{sub:nfac}  for latent factor selection, common $K$ and specific $J_s$,  proved to be extremely robust respect to the choice of this threshold. Conclusion with a threshold of 0.1 was the same. We choose a lower value as are more concerned to lose important  shared biological factors than to include additional shared factors.
\begin{figure}[!t]
\centering
\resizebox{1\textwidth }{!}{ %
\includegraphics[width=1.6cm]{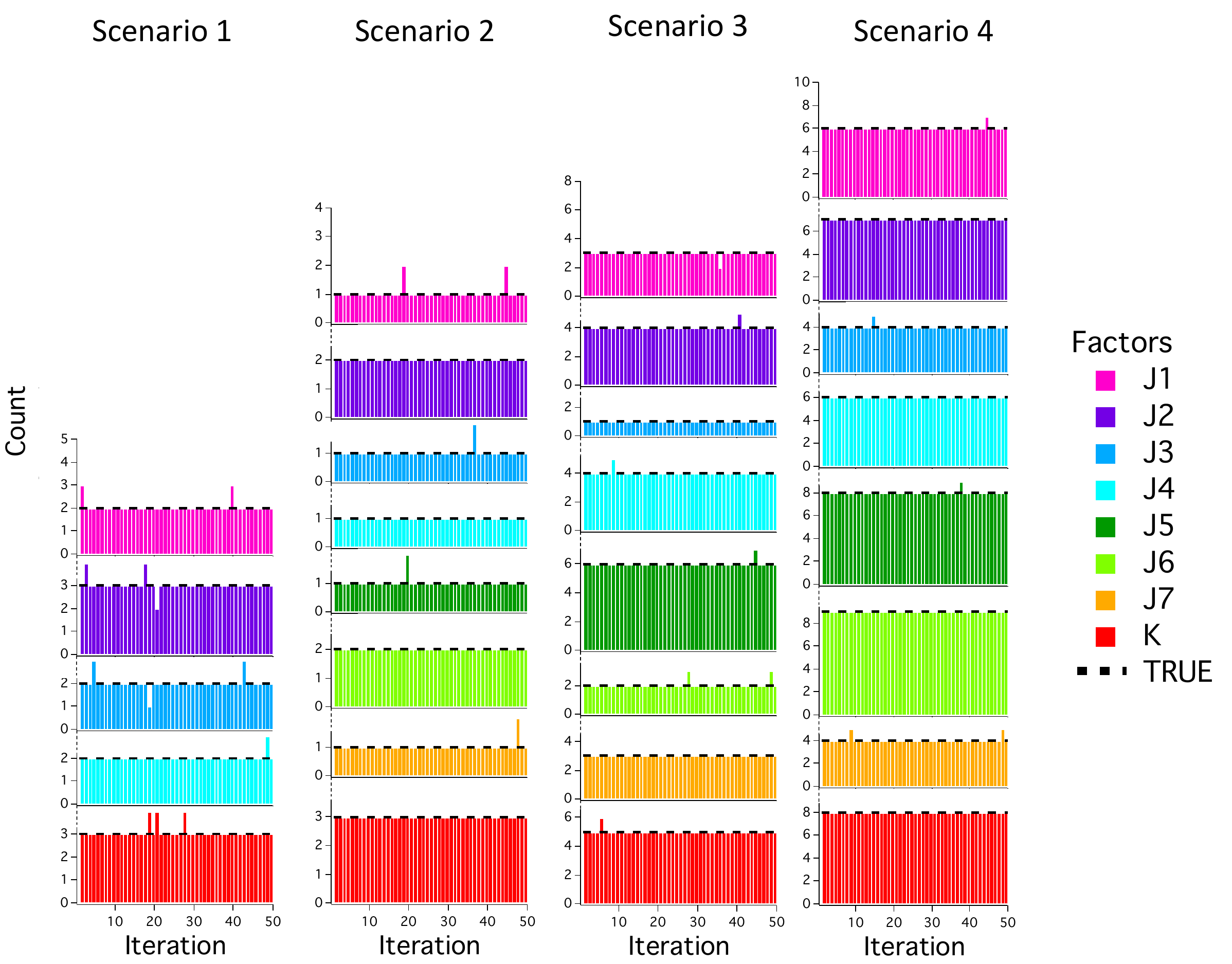}
}
\caption{\it Dimensions of shared and study specific factors in the four scenarios. Model selection procedure for the shared $K$ and the study-specific $J_s$ latent dimension via SVD of $\bsS_\Phi$ and $\bsS_{\Lambda_s}$. The true dimensions are visualized by the dashed lines.}
\label{fig: scelta}
\end{figure}
Figure~\ref{fig: scelta} shows the results obtained by fitting the model for 50 different data sets generated from the BMSFA with $K = 3$ in the four different scenarios.  The vertical lines show the 50 estimated latent dimensions in each data set. Our method consistently selects the right dimensions for both the shared and the study-specific factors.  

The simulation analysis highlights the merit of our method in a variety of scenarios, with improved performance over FA in terms of covariance matrices estimation in multi-study settings, estimation of the reproducible signal across studies, and identification issue for the factor loading matrix.

%-------------------------------------------------------------------------
\section{Breast Cancer Case Study}\label{sec:CS}

The aim of this analysis is to identify shared common factors describing the common correlation structure across the 7 breast cancer microarray studies listed in Table~\ref{tab: dataset}. Recovering shared gene co-expression patterns from different high-throughput studies is important to identify replicable genetic regulation. This case study considers a relatively well understood area of cancer biology and provides a realistic positive control for the BMSFA methodology.

\begin{figure}[!t]
%\includegraphics[height=8cm, width=8cm]{stime_insieme.pdf}
%, trim={1cm 5.2cm 0cm 5.4cm},clip
\centering
 \includegraphics[height=14.26cm, trim={0cm 5.2cm 0cm 5.4cm},clip]{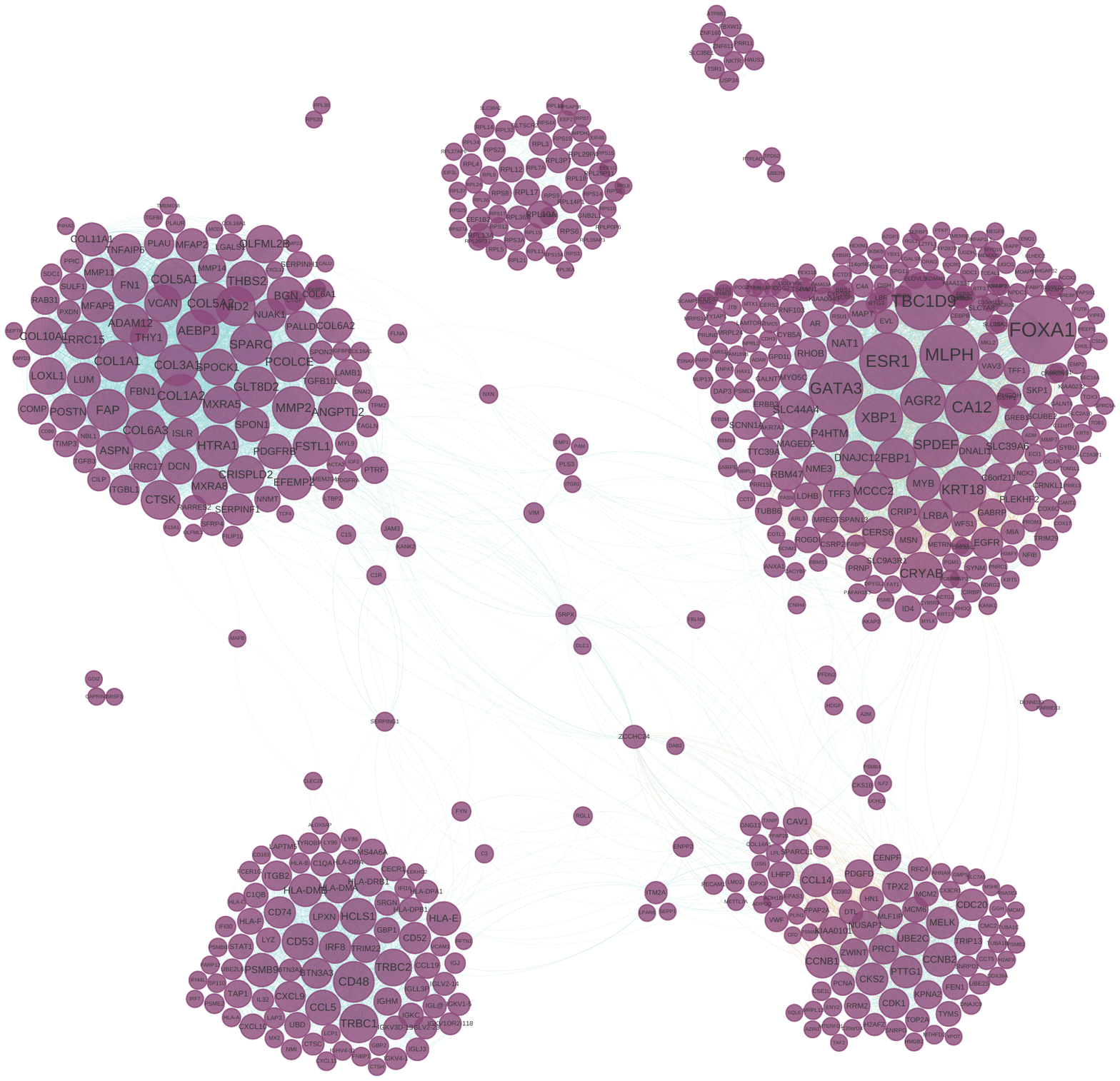}
 \caption{\it \small Shared gene co-expression network across the 7  studies of Table~\ref{tab: dataset}. We include edges between two genes if the corresponding element in the shared part of the covariance matrix is greater than 0.5. Edges in blue (orange) represent positive (negative) associations. 
}
\label{fig: cluster}
\end{figure}

We consider genes measured in all studies and remove the $50\%$ of genes with the lowest variance. We use the prior of Table~\ref{tab:priSim}.   
Our method chooses a shared latent dimension of $K=8$, through the SVD of $\bsS_\Phi$. We first summarize and visualize the shared co-expression patterns via a co-expression network (Figure~\ref{fig: cluster}) built on $\bsS_{\Phi}$, and thus representing all studies. A gene co-expression network is an undirected graph. Each node corresponds to a gene and each edge represents a high co-expression between genes. The importance of genes in a cluster is represented by the node size.

Our analysis identifies five larger clusters. Co-expressed genes tend to be members of the same, highly plausible, biological pathways. All clusters are associated with biological processes known for explaining heterogeneity of expression across breast cancers, lending credibility to BMSFA.   
The first cluster is driven by expression of the estrogen receptor (ESR1), which historically is one of the earliest cancer biomarkers to have been discovered, and plays a crucial role in the biology and treatment of breast cancer \citep{jordan2007, robinson2013}. High dimensional expression pattern are found in \cite{sorlie2001}. Many studies have shown the relation of ESR1 with growth of cancer \citep{osborne2001, iorio2005, toy2013}. Levels of ESR1 expression are associated with different outcomes \citep{ross2012, theodorou2013}. Three other genes stand out: GATA3, XBP1 and FOXA1. These are ESR1-cooperating transcription factors altered in breast tumors \citep{lacroix2004, theodorou2013}.  In breast cancer cell, many studies revealed strong and positive association of GATA3, XBP1 and FOXA1 with ESR1 \citep{hoch1999, sotiriou2003, sorlie2003, lacroix2004, lai2013,  theodorou2013}.
The second cluster is related to the cell cycle. One of the most important genes in this cluster is CCNB1, which encodes cyclin B. Cyclins are prime cell cycle regulators. Many analyses found a common pattern of overexpression of the mitotic cyclins A and B and their dependent kinase in the tumor cell of breast cancer \citep{keyomarsi1993, lin2000, basso2002}. Two other important genes in this cluster are CDK1, a  kinase dependent on cyclins, and CDC20, a gene related to the metaphase and anaphase of cell cycle.
All genes in the third cluster are related to regulation of the immune response. The CD genes are important for the immune system pathway and the HLA genes are a crucial element for immune function.
The fourth cluster includes several genes expressed by the connective tissue, including collagen genes (COL1A1, COL1A2, COL3A1, COL5A1, COL5A2, COL10A1, COL11A1), previously associated with stromal cells \citep{ross2000, ioachim2002}. Of note are also ADAM, a protease related to the degradation of the connective tissue, and smooth muscle cell marker TAGLN, also previously found to play a role in breast cancer.
Finally, all the RP genes in the fifth cluster codify the ribosome, which synthesizes proteins. Dysregulation of Ribosome function is related to tumor progression in breast cancer \citep{belin2009}.
 
\begin{figure}[!t]
\centering
 \includegraphics[height=5.26cm]{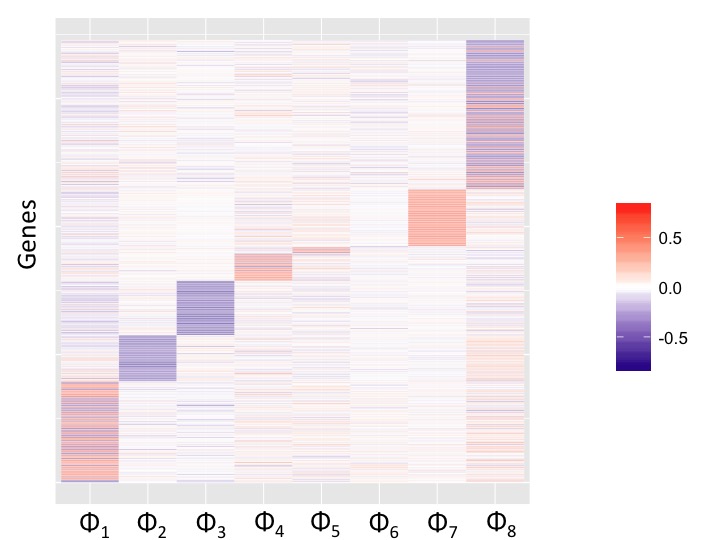}
 \caption{\it \small Heatmap of the estimated shared factor loadings obtained with BMSFA across the 7 studies in Table~\ref{tab: dataset}. We only show common factor loadings $\geq  0.5$.}
\label{fig: heatmap}
\end{figure}

To further explore the patterns found in the shared gene co-expression network, we estimate the shared factor loadings after the post-process procrustes algorithm. 
The heatmap in Figure~\ref{fig: heatmap} depicts the
estimates of the shared factor loadings
that can be identified reproducibly across the studies. 
To extract biological insight from the shared factors, we explore whether specific gene sets are enriched among the loadings using Gene Set Enrichment
Analysis (GSEA) \citep{Mootha2003}. We used the package \texttt{RTopper} in \texttt{R} in \texttt{Bioconductor}, following the method of \citet{tyekucheva2011} and considering all the gene sets representing pathways from {\tt reactome.org}. The resulting analysis shows concordant results with the pathways obtained with the shared gene co-expression network, further suggesting that we identify genuine biological signal. The first shared factor is significantly enriched with the ``Cell communication" and ``Cell cycle" pathways. The second factor is associated with the Immune system pathway and all the sub-pathway included in it, namely the ``Adaptive Immune System", ``Innate Immune System" and ``Cytokine Signaling in Immune System". Factor 3 shows a significant association with cell cycle, namely with the pathway ``Cell cycle", ``Cell cycle mitotic", ``Cell cycle checkpoints", ``Regulation of mitotic cell cycle". The shared 5, 6 and 7 factors have protein production   ``Transport of ribonucleoproteins into the host nucleus", ``Protein folding", ``Mitochondrial protein import", ``Metabolism of proteins", ``NRIF signals cell death from the nucleus". Finally, factor 8 is related to the ER pathway, ``ER phagosome pathway", ``Interferon signaling", and ``Interferon alpha beta signaling".

\begin{figure}[!t]
	\centering
	\includegraphics[width=5.2cm]{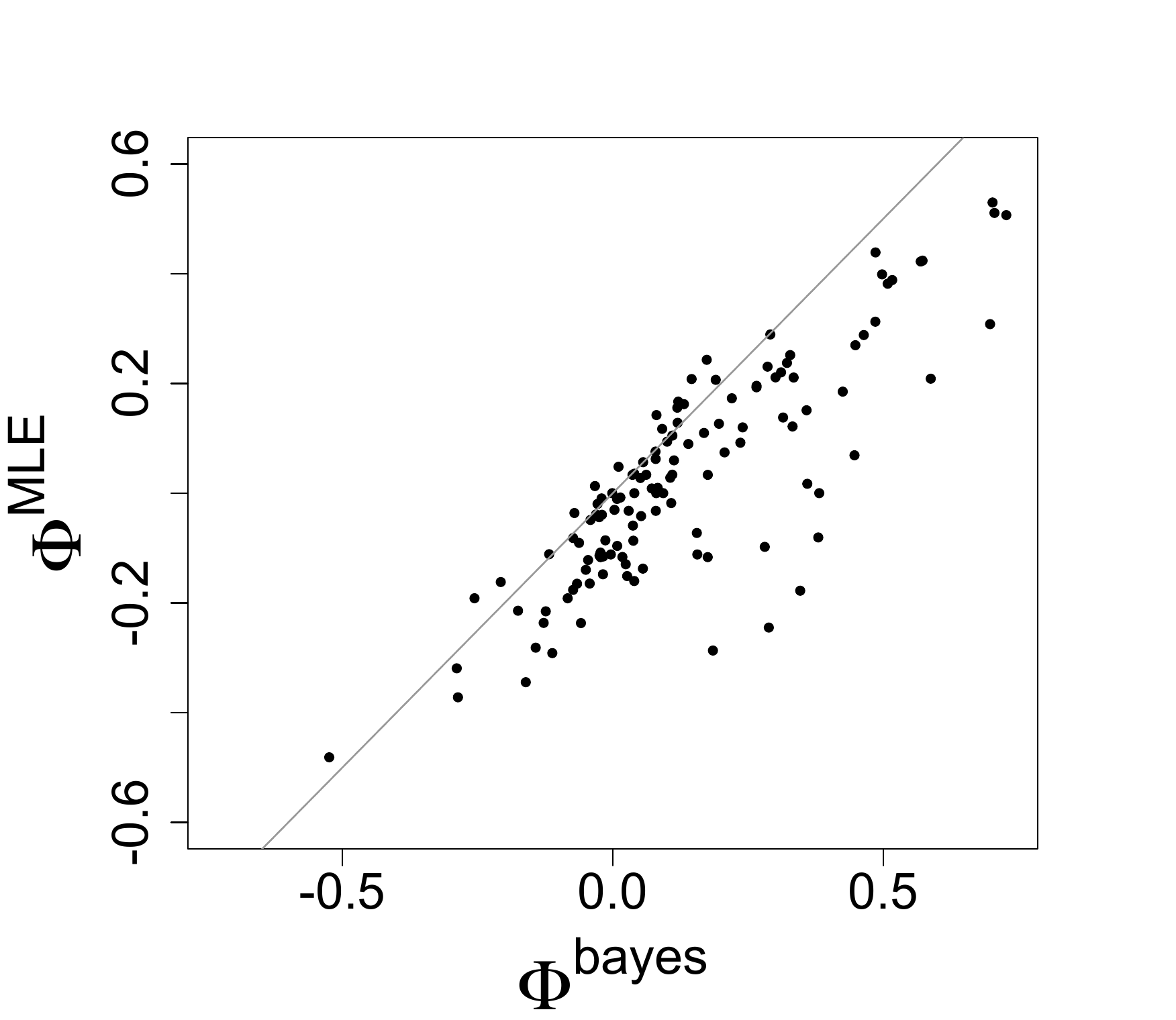}
	\includegraphics[width=5.2cm]{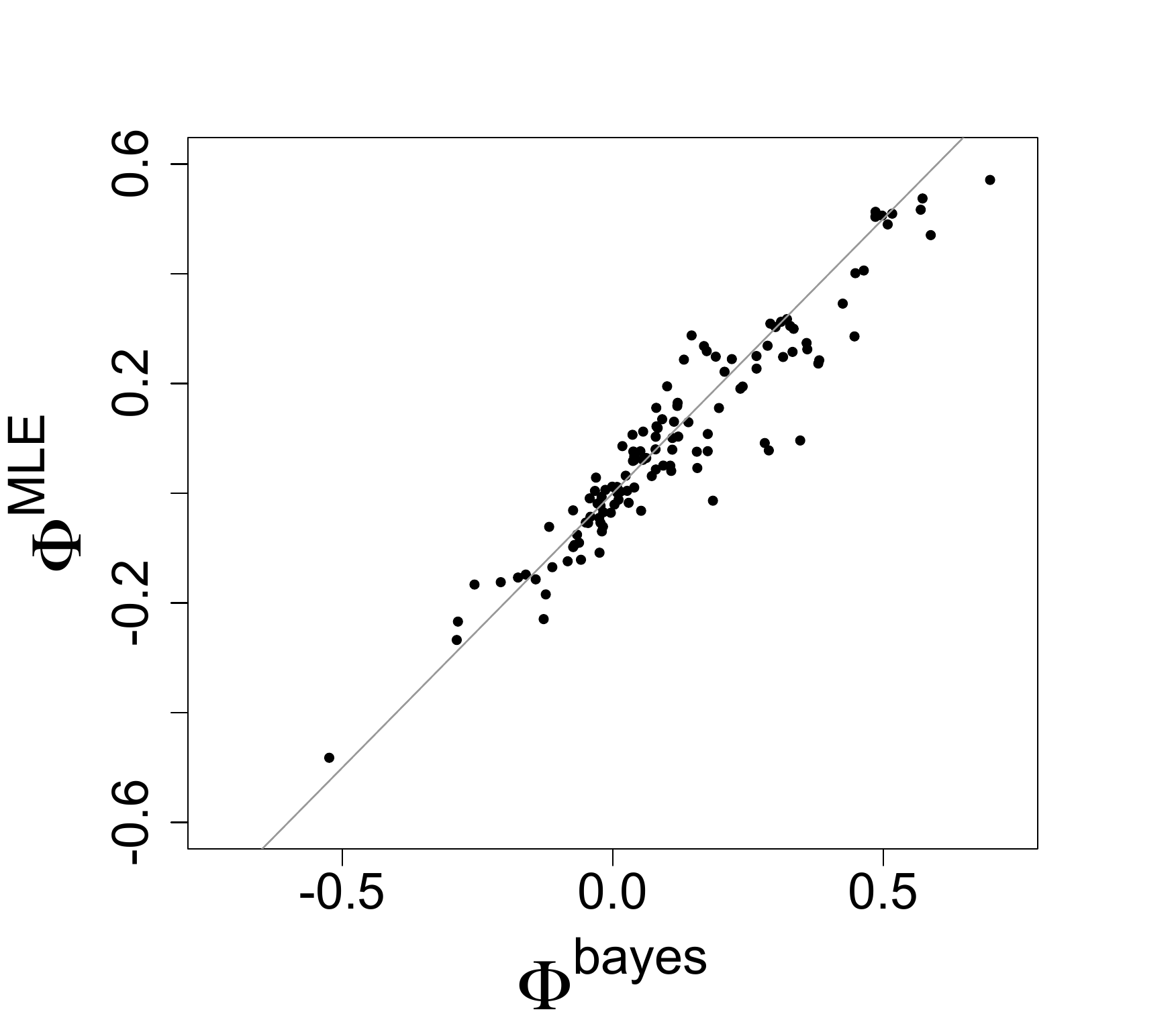}
	\caption{\it Left: Comparison between the first two estimated common factor loadings with MLE and BMSFA. Right: Comparison between the first two estimated factor loadings with MLE, followed by varimax rotation, and BMSFA.}
	\label{fig: MLEbayes}
\end{figure}

%%%% Shrinkage comparison
An important feature of BMSFA in this case study is regularization of the common factor loadings. To illustrate this in more detail, we conclude this section comparing BMSFA to the MSFA which uses MLE for parameter estimation. The data consists of 63 genes in the Immune System Pathway. Their loadings are compared in Figure~\ref{fig: MLEbayes}.
BMSFA regularizes common factor loadings by shrinking small and moderate MLE loadings to zero while systematically amplifying larger MLE loadings (Figure~\ref{fig: MLEbayes}, left panel). This regularization behavior is somewhat unique to this setting, as it is far more common for regularization to only result in shrinkage. Here, the prior helps the posterior perform a factor rotation method which results in more sparse factors.  To further illustrate we rotate the loadings obtained with the MLE using the varimax rotation  \citep{kaiser1958} and we compare it with the BMSFA (Figure~\ref{fig: MLEbayes} right panel). The BMSFA loadings are far more similar to the varimax rotated MLE (correlation 0.97) than the original MLE (correlation 0.7).

\section{Discussion}
In this paper we propose a general Bayesian framework for the unsupervised analysis of high-dimensional biological data across multiple studies, building upon~\cite{DV2016}. We address the unmet need to rigorously model replicable signal across studies, while at the same time capturing study-specific variation. Our approach is not limited by $P\ll n$ and, in addition to replicability, shows considerable promise in modeling sparsity and enhancing interpretability via rotation-like shrinkage. Building on \cite{bhattacharya2011} we propose a computationally efficient MCMC algorithm.

The work in this paper is motivated by identifying replicable signal in unsupervised genomic  applications. The results of our analysis of breast cancer gene expression across seven studies identified shared gene patterns, which we also represented via clusters in a co-expression network. Both factors and clusters are clearly related to well-known breast cancer pathways. Our analytic tools allows investigators to focus on the replicable shared signal, after properly accounting for, and separating, the influence of study-specific variation. While we focused on the shared signal, study-specific loadings can also be examined directly.

BMSFA may have broad applicability in a wide variety of genomic platforms, including microarrays,
RNA-seq, SNP-based genome-wide association studies, proteomics, metabolomics, and epigenomics. Relevance is also immediate in other fields of biomedical
research, such as those generated by exposome studies, Electronic Medical Record (EMR), or 
high dimensional epidemiological data. In dietary pattern analysis, it is important to find replicable dietary patterns in different populations \citep{castello2016}. 
Our analysis could be applied to check if there are shared dietary patterns across different populations and to detect the study-specific dietary patterns of a particular population. In this field generally, it is common to apply a varimax rotation to factor loading matrix, for a better interpretation. Specifically, the interpretation of a factor relies on loadings. The interpretation of the model is simplified if more of the loadings are shrunk towards zero and the factor is defined by few large loadings. In the frequentist analysis, this is possible by rotation methods, such as varimax.  In our representation, the BMSFA embeds this step giving an immediate representation of the two sparse factor loading matrices through the shrinkage prior, as shown in Section~\ref{sec:CS}. 

%Alternative proposals for the analysis of multiple studies include the Multiple Co-inertia analysis (MCIA) \citep{dray2003, meng2014} and Multiple Factor analysis (MFA) \citep{abdi2013}. However, for our goals, both MCIA and MFA are limited. First, MCIA and MFA are focused on analyzing only the common structure and fail to consider the study-specific part. Instead, our method estimates both common and study-specific components. Second, MCIA and MFA operate stage-wise, decomposing each matrix separately, while our study analyzes the datasets jointly in one single step. 
Our Bayesian non-parametric approach offers more flexibility in the choice of the dimensionality of shared latent factors. Moreover, we provide shrinkage of the latent factor loadings, enhancing the role of the variables that are most important in each factor.
%Our approach is really adaptable to researchers need in choosing either parametric or non-parametric specification for various model components, which is a key consideration especially in high-throughput data analysis and in the $p \gg n$ context.
 
To address the choice of model dimension, we developed, building on \cite{bhattacharya2011}, a practical procedure based on separate SVD of the shared covariance part and the study-specific covariance parts. The choice of the number of factors remains an important open problem. The most common method for choosing latent dimension fits the factor model for different choices of $K$ and compares them using selection criteria such as BIC. This approach presents many problems especially in a $p \gg n$ setting where MLE is not duable.  \cite{lopes2004} proposed a reversible jump MCMC to estimate the number of factors in standard FA, but this method is also often computationally intensive.  \citet{bhattacharya2011} developed an interesting adaptive scheme that dynamically changes the dimension of the latent factors as the Gibbs sampling progresses.  In our approach, we develop a practical approach where we have a balance between retaining important factors and removing the redundant ones.
 
We also address identification. Identifiability remains a challenge in standard FA. In the Bayesian approach, constraints were proposed to tackle this issue, such as that of a block lower triangular matrix \citep{lopes2004, carvalho2008}. As \cite{carvalho2008} noticed, in this constraint different ordering of variables could lead to different conclusions. In our work, we adopt a procrustes algorithm and demonstrate through a series of simulation analyses that this method applied to the BMSFA is effective. \citet{rovckova2016} solves this problems in a Bayesian context by rotating the factor loadings matrix with the varimax rotation \citep{kaiser1958}. We also compared the BMSFA estimates after the procrustes algorithm with the MLE after rotating the common factor loadings. The resulting analysis are close, demonstrating that the prior we adopt works similarly to a rotation.

We hope BMSFA will encourage joint analyses of multiple high-throughput studies in biology, and contribute to alleviating the current challenges in replicability of unsupervised analyses in this fields and across data science.

\clearpage
\bibliographystyle{apa}
\bibliography{RDVbayes,gpRefs}
\end{document}